\documentclass{article}[12pt]
\usepackage{epsfig}

\usepackage{amssymb}
\usepackage{longtable}
\usepackage{url}
\usepackage{amsmath}
\usepackage{amsfonts}
\usepackage{amsthm}

\newcommand{\superscript}[1]{\ensuremath{^{\textrm{#1}}}}

\urldef{\mailmd}\path|mpdonato@students.latrobe.edu.au|

\begin{document}
\label{firstpage}

\title{EcoHomeHelper: An Expert System to Empower End-Users in Climate Change Action }

\author{Matt Donato and Seng W. Loke\\
       Department of Computer Science and Computer Engineering\\
La Trobe University, VIC 3086, Australia\\
\tt{m.donato@latrobe.edu.au, s.loke@latrobe.edu.au}}
 

\maketitle

\begin{abstract}
Climate change has been a popular topic for a number of years now.  Computer Science has contributed to aiding humanity in reducing energy requirements and consequently global warming.  Much of this work is through calculators which determine a user's carbon footprint.  However there are no expert systems which can offer advice in an efficient and time saving way.  There are many publications which do offer advice on reducing greenhouse gas (GHG) emissions but to find the advice the reader seeks will involve reading a lot of irrelevant material.  This work built an expert system (which we call EcoHomeHelper) and attempted to show that it is useful in changing people's behaviour with respect to their GHG emissions and that they will be able to find the information in a more efficient manner.  Twelve participants were used.  Seven of which used the program and five who read and attempted to find advice by reading from a list.   The application itself has current implementations and the concept further developed, has applications for the future.
\end{abstract}

\section{Introduction}
\label{intro}

One of the greatest challenges to humanity in the 21\superscript{th} century is global warming \cite{reay05}. An increase in global temperature of between 2 - 3\superscript{o}C could be considered as causing irrevocable damage \cite{ipcc01}. However, an increase of just 1\superscript{o}C to some may be a closer estimation \cite{hanson07}. The potential effects for this planet are both dramatic and potentially devastating.  ``Climate change may be due to natural internal processes or external forcings, or to persistent anthropogenic changes in the composition of the atmosphere or in land use.''  While it is beyond the scope of the current work to discuss this extensive work in detail, it is suffice to note that the IPCC has found an increase of 0.6\superscript{o}C globally of average surface temperature during the 20\superscript{th} century.  Sea levels rose by 0.1 to 0.2 meters in that century.  A continuing increase of greenhouse gases is a direct result of human activity and consequently there is powerful evidence that in the last 50 years global warming is attributable to anthropogenic activity.

This paper will focus on helping end users change their behaviour with regards to climate change.  The paper proposes and develops an expert system in Prolog which will give advice to users regarding their GHG emission in the residential sector.

The rest of this paper is organized as follows.  Section~\ref{related} will look at related work and predominantly at expert systems used to aid people change their energy requirements.  In section~\ref{helper}, EcoHomeHelper is presented along with the problem description and its solution.  Section~\ref{application} depicits the code used to develop the application.  Section~\ref{usability} gives the method and results from the conducted study with section~\ref{discussion} discussing the findings.  Finally, in section~\ref{conclusions} a conclusion to the study is given followed by possible future work arising from this study.

\section{Related Work}
\label{related}

This section will discuss work related to computer science and the monitoring of GHG emissions in the residential area.  Weight is give to the only study found were an expert system was employed.

\subsection{Computer Applications and Persuasion}
\label{persuasion}

It may appear strange to some readers to be considering Social Psychology issues when attempting Computer Science research.  However, this author believes the union is at times inevitable and indeed unavoidable when building an interactive application offering advise.  In discussing Human-Computer Interaction (HCI), from \cite{bickmore05} we can confidently conclude that people do form social-emotional relationships with such things as jewelry, clothing and other non humanoid physical or nonphysical forms.  The authors further suggest that people can respond to computers in social ways even though they may be unconscious of this behaviour and consequently form relationships with computers.  The above article continues to argue that due to this social-emotional bonding with the agent, in this case the application, computers can play a role in aiding people to change their behaviour.  In other words, computer applications can play a role in persuasion.

\subsection{Expert Systems in GHG Emission Monitoring}
\label{expertSystems}

A generous and robust search on Google and Google Scholar and further using the resources of the La Trobe University library has yielded only one expert system which deals with climate change \cite{benders06}.  There are many calculators available on the web which deal with users calculating their GHG emissions with respect to energy usage in the home \cite{epa08} \cite{susvicgwp} \cite{gore06} \cite{rose06}.  There is even a calculator which determines the carbon stored in a tree \cite{CRC08}.  It beggars belief that with such an important topic as climate change and it's anthropogenic influence and the far reaching impact of computer science in western society, that no one has developed an expert system to aid people fight global warming.  This section will discuss that article \cite{benders06} and its use of the expert system.

The authors attempted to measure energy reduction in households in the Netherlands by participants using the tool.  The researchers wanted to evaluate the tool and determine its capability to inspire participants to reduce their total energy requirement.  

Further, they asked if the tool was able to address indirect energy requirements? Direct energy requirements for households were defined as the consumption of electricity, natural gas and motor fuel.  Indirect energy on the other hand is the production, distribution and waste disposal of consumer goods and services.  The tool is described as being web based, interactive and scalable.  They argued that the tool was personalized because participants were asked to choose from a list of appliances (for example clothes dryer) and a negative response would result in no questions presented to them regarding that item.  The researchers stated that participants could access a help desk either by mail or phone.  It was designed so participants could complete the questionnaire within 45 minutes, and energy reductions were shown as a percentage of total energy needs.  These latter design issues were to make the tool user friendly, and consequently participant dropout would be minimal and problems would be manageable.

Participants were selected by either responding to a local newspaper advertisement or by direct mail.  A total of 5,000 letters were randomly sent to households.  They had 443 respondents.  However, only 347 commenced the experiment.  The participants were placed into either a control or experimental group.  The former received no energy saving option, personalized reduction options nor feedback during the experiment.  However, they did receive personalized reduction options after the experiment.  The latter group received information regarding energy saving and personalized reduction options and personalized feedback.  Three surveys were given from October 2002 to February 2003.  The control only received the first and last questionnaire.  The aim of the first questionnaire was to provide an energy reduction option list generated by the expert system.  The questions centered on electricity and natural gas needs in the household.  This determined the total energy requirement for the household.  The second survey asked participants about any changes in the household i.e. appliances, cars.  The third questionnaire was concerned with questions about holidays and providing feedback regarding energy reduction compared to that at the beginning of the experiment.

Of the 347 participants 190 completed the experiment.  The exclusion of the 157 participants were due to either participants having no time or their computer being broken, calculations were performed incorrectly, the household composition had changed or essential answers were not provided.  

The results show a reduction of direct energy requirements for the experimental group compared to the control group.  There was no statistically significant difference between the two groups with indirect energy reduction.  The study used an ANOVA test.

The researchers concluded that while energy conservation was statistically significant between the two groups, total energy reduction was not significantly different.  This was largely due to energy requirements when participants were on holidays.  The authors admit that the tool did not reach the public as in a total only 4\% of 5,000 invitations completed the experiment.  They believe that the tool is able to measure indirect energy needs but scalable only for direct energy needs.

A weakness of the above paper is that no adequate discussion of the expert system is discussed.  The reader is left wondering if the tool was a shell with data added, or if it was written using a particular programming language, Prolog or LISP, for example.  The sample size was small, but the authors do suggest larger sample sizes are needed in future studies.  Another weakness was the length of time required to complete the questionnaire.  45 minutes was too long and it is not surprising that ``no time'' was a contributing factor to the participant dropout rate.

\section{EcoHomeHelper}
\label{helper}

Section~\ref{expertSystems} outlined the only expert system found by this study.  It presented the difficulties with that system and indicated the need for such a system to be developed.  This section will present the problem description and a solution to building an expert system which will give people advice in reducing their carbon footprint.

\subsection{Problem Description}
\label{problem}

With any expert system the initial problem is that some expertise in the area is required.  This can be achieved by consulting an expert in that domain but for small programs, the programmer can assume that role \cite{neg02}.  There are many publications available which provide comprehensive advice or tips to the public \cite{gore06} \cite{blackburn07} \cite{coolit08}.  The difficulty lies in presenting the advice in a way the user can access it.  For example, the reader will need to sift through a lot of information which may be irrelevant.  Finding particular information may be quite time consuming.  Some publications \cite{blackburn07} not only offer tips on climate change but also on ``green living''.

Approximately 100 tips or advice in total on how to reduce energy requirements in the home where taken from \cite{gore06} \cite{blackburn07} \cite{coolit08}.  These bits of advice are organized into different categories, Lighting, Transport or Buying a home, for instance. Textbooks on the subject will organize the information in this way.  Even so, as mentioned earlier, the reader must read a lot of irrelevant information.  However, for an expert system to access knowledge in only one way is not adequate and other ways to retrieve the information are required.  Different bits of advice may relate to more than one topic.  For instance, if one wishes to know at what temperature to set a thermostat, the information should be accessible not only by inquiring about thermostats but also by accessing appliances which use thermostats like fridges or clothes dryers.

\subsection{Problem Solution}
\label{solution}

The amount of information to collate is vast as indicated in Section~\ref{problem}.  In this section a solution to organizing the data is presented.  As with the texts \cite{gore06} \cite{coolit08} \cite{blackburn07} from which the data came from, the information was categorized by an aspect of the home.  In the knowledge base this was called \texttt{area}. Given the vastness of information the tips were broken down into \texttt{advice} and \texttt{rationale}.  Dividing the data in this way gives us an enormous advantage.  Some rationale includes facts on GHG emissions. For example, the snippet below gives a quantifiable amount of GHG emissions.  This has allowed this author to add a tag \texttt{ghg} which in turn allows the user to access specifics facts or values for GHG emissions as opposed to general information:
\begin{verbatim}
advice(area(Hot Water Systems),
    stage(Buying),
    type(Type of Hot Water System),
    ghg(Greenhouse Gas Emissions Facts),
    theAdvice(Don't use a hot water system with a continuous pilot.),
    rationale(This can save $40 and 200kg of GHGs per year.)).
\end{verbatim}
The above advice can be read in this way.  Firstly, \texttt{advice} refers to a fact in the knowledge base.  Encapsulated in this are \texttt{area(Hot Water System)} which refers to the aspect of the home the use user wants to inquire about, \texttt{stage(Buying)}, this defines if the user is interested in buying a product and would like to know how to save on energy requirements when purchasing that commodity.  In the above example, a Hot Water System.  The \texttt{type} tag binds the \texttt{area} with a specific product or appliance.  Continuing to use the above fact \texttt{type} refers to features of Hot Water Systems.  Other \texttt{type} for Hot Water Systems is \texttt{Showers and Taps}.

As previously mentioned, the final two tags are \texttt{advice} and \texttt{rationale}.  The former refers to the advice sought by the user and the latter gives an explanation for that advice.  This was done as there was too much information to be placed in one section.

There was hope that in using multiple tags, the same facts were accessible from different pathways.  For example, the user could access information about living closer to work and schools via selecting \texttt{Transport} or the option \texttt{Buying a Home}.  A further need for using multiple tags arose when considering what may be useful for the user.  It presented a solution to present the user with menus.  The user could select an option relevant to his or her needs or interest.  The tags and their effectiveness are better exemplified in the discussion of the application.  The next section is devoted to presenting the application.

\section{The Application}
\label{application}

The Prolog language arose from the work of Robert Kowaiski and Alain Colmerauer during the early 1970s \cite{roth02}.  As a rule based language, Prolog provides a flexibility and is suited to variable combination of rules \cite{cook90}. This type of language is suitable when executing the solution outlined in section~\ref{helper}.  There are number of applications written in Prolog.  Specifically SWI-Prolog.  Applications range from a Biological knowledge integration application (Blipkit), to a multiuser deductive database used for meta modeling and engineering of customized modeling languages (ConceptBase).  There are many other examples of SWI-Prolog applications which the reader can find at \cite{apps08}.  In short, this author attempted to find a solution to the problem of presenting a vast amount of information which could be programmed using a flexible language.  Prolog presents such a language.

\subsection{The Code}
\label{code}

The first predicate used is to display the Main Menu.  The options here are hard coded.  The predicates which display the menus are code modified from AMZI Logic Explorer Tutorial \cite{amzi08}.  All other predicates produce results dynamically.  It produces menu options from the knowledge base therefore when new facts need to be added or if the programmer wishes to remove some facts, only the knowledge base need be changed.  The predicates use the inbuilt functions findall and sort.  The findall predicate uses the variable entered by the user to find only those specific facts from the database.  The sort function is used because of its ability to delete duplicate information.  This gives the application the flexibility and efficiency to create the next menu options.  The following code illustrates this.
\begin{verbatim}
areaMenu:-
    nl,
    write('Aspect of Home Menu'), nl,
    findall(Area, advice(area(Area), _, _, _, _, _), TheArea),
    sort(TheArea, AreaList ),
    menuask(below, Picked, AreaList),
    stageMenu(Picked).
\end{verbatim}
As we can see a list is created in findall.  Duplicates are eliminated using sort.  This final list, in our example \texttt{AreaList} is passed down to the \texttt{menuask/3} rule along with the variable \texttt{Picked} which will hold the user's choice.  The \texttt{menuask/3} generates the menu and the choice, \texttt{Picked} is then passed to the next predicate for more menu options.  The structure of all menu predicates are the same, except they hold the value of the user's previous choice, with one exceptions.  The last menu predicate \texttt{ghgMenu/3} calls another predicate \texttt{finalAdvice/4}.  This latter rule collects all the variable choices made by the user, creates a list and then calls the predicate responsibly for displaying the output.  The following shows the structure of \texttt{findAdvice/4}.
\begin{verbatim}
finalAdvice(S, A, T, G):-
    findall(advice(area(A), stage(S), type(T), ghg(G), Ad, R), advice(area(A), 
    stage(S), type(T), ghg(G), Ad, R), FinalAdvice),
    giveAdvice(FinalAdvice).
\end{verbatim}
Finally, \texttt{finalAdvice/4} calls \texttt{giveAdvice/1} which generates the desired advice and then displays that advice to the screen.  The rule is quite simple as shown below.
\begin{verbatim}
giveAdvice([]).
giveAdvice([Head | Tail]):-
    Head = (advice(_, _, _, _, theAdvice(Advice), rationale(Rationale))),
    nl,
    write('Advice : '),
    write(Advice),
    nl,
    write('Rationale : '),
    write(Rationale),
    nl,
    giveAdvice(Tail).
\end{verbatim}
There is also a rule which checks for errors.  If the user enters a zero or a number higher than the total number of options generated then an error message is called and the menu is displayed again.  Otherwise, the next phase in the application is executed.  The following code shows this.
\begin{verbatim}
checkOption(Num, Val, Alist):-
    length(Alist, Len),
    ( ( Num < 1) ->
    nl,
    write('***** ERROR!  Cannot Have Zero Options\! ******'),
    nl,
    menuask(below, Val, Alist);
    ( (Len < Num) ->
    nl,
    write('***** ERROR!  Number Entered is Greater than number of Options\! *****'),
    nl,
    menuask(below, Val, Alist));
    ( (Len >= Num) ->
    nl,
    pick_menu(Num, Ans, Alist),
    Val = Ans)).
\end{verbatim}
This author has attempted to present the code used in the application.  The most important aspect is that all options and advice are generated dynamically.  Changes need only be make to the knowledge base.  The next part of this work is to focus on the implementation of the application.

\section{Usability and Testing Aims}
\label{usability}

We now present results from a questionnaire completed by participants.  It will discuss the participants, the method of the study and results.

\subsection{Participants}
\label{participants}

A total of twelve participants were used in this study.  They were randomly selected from university students, a social club and nurses working in a mental health inpatient unit.

\subsection{Method}
\label{method}
   
The participants were divided into two groups.  All were asked to give some basic demographics such as age, computer competency and whether they were conservationists.  Seven were asked to use EcoHomeHelper, while five were asked to read the advice generated by EcoHomeHelper from print.  All participants were given the same four scenarios.  The first scenario involved the user wanting advice on buying a new fridge.  The second asked to find advice to reduce GHG emissions with respect to computers, stereos and TVs in the home.  The third asked the user to find an alternative to driving the car everywhere.  Finally, the fourth scenario directed the user to get advice on alternatives to using the air conditioner in summer and the heater in winter.  Participants who used EcoHomeHelper were asked to write the options they chose, and if they could not find the advice they wanted and had to restart the program, to write \texttt{restart}.  Participants were give five minutes to \texttt{play around} with EcoHomeHelper to become familiar with the application.  Participants were timed to see how long it took to complete the four scenarios.

After completing the scenarios.  Participants who used EcoHomeHelper were asked to complete five questions using the Likert Scale \cite{brit08}.  The questions asked if the options of the application were easy to follow.  They found the advice the wanted easily.  The advice was helpful.  If the advice was helpful to change their behaviour with respect to reducing GHG emissions.  Finally, whether they would have such an application on their desktop or mobile phone.

Lastly, those who used EcoHomeHelper were asked to give comments on whether they liked the application, how it could be improved and if such an application would be useful on a mobile phone or PDA.

\subsection{Results}
\label{results}

After gathering the participants' responses, some statistical analysis was performed on the data.  A detailed picture is given in table~\ref{table:time}.  The first column \texttt{mode of Testing} shows whether it was reading from a list or using the program.  The next column \texttt{N} gives the sample size for each group.  The remaining columns show the \texttt{Mean}, \texttt{Standard Deviation} and \texttt{Standard Error of Mean} for each group.  The mean time for completing the tasks were 14.4 minutes and 9.86 minutes for those who read from the list and those who used the program respectively.

\begin{center}
\begin{longtable}{| l | l | l | l | l |}
\caption{Time Completion} \label{table:time} \\
\hline 
\multicolumn{1}{|c|}{\textbf{Mode of Testing}} & 
\multicolumn{1}{c|}{\textbf{N}} &
\multicolumn{1}{c|}{\textbf{Mean}} &
\multicolumn{1}{c|}{\textbf{Std. Deviation}} &
\multicolumn{1}{c|}{\textbf{Std. Error of Mean}} 
\\ \hline 
\endfirsthead
\multicolumn{5}{c}
{{\bfseries \tablename\ \thetable{} -- continued from previous page}} \\
\hline 
\multicolumn{1}{|c|}{\textbf{Mode of Testing}} & 
\multicolumn{1}{c|}{\textbf{N}} &
\multicolumn{1}{c|}{\textbf{Mean}} &
\multicolumn{1}{c|}{\textbf{Std. Deviation}} &
\multicolumn{1}{c|}{\textbf{Std. Error of Mean}} \\
\hline
\endhead
\hline \multicolumn{5}{|r|}{{Continued on next page}} \\ \hline
\endfoot
\hline \hline
\endlastfoot
\textbf{Reading From List} & 5 & 14.4 & 4.336 & 1.939 \\
\textbf{Use Program} & 7 & 9.86 & 3.132 & 1.184 \\
\hline
\end{longtable}
\end{center}

The exact values for the likert scale data are given in Table~\ref{table:likert}.  The scores on each of the items were added to give a total score and then a mean found.  It shows the the questions relating to \texttt{Advice was Easy to Follow} and \texttt{Advice Was Helpful} were the highest scoring. 
\begin{center}
\begin{longtable}{| l | p{0.3cm} | p{0.1cm} | p{0.1cm} | p{0.5cm} | p{.7cm} |}
\caption{Exact Values From Likert Scale Questions} \label{table:likert} \\
\hline 
\multicolumn{1}{|c|}{\textbf{}} & 
\multicolumn{1}{c|}{\textbf{N}} &
\multicolumn{1}{c|}{\textbf{Min}} &
\multicolumn{1}{c|}{\textbf{Max}} &
\multicolumn{1}{c|}{\textbf{Mean}} &
\multicolumn{1}{c|}{\textbf{{\footnotesize Std. Deviation}}} \\ 
\hline 
\endfirsthead
\multicolumn{6}{c}
{{\bfseries \tablename\ \thetable{} -- continued from previous page}} \\
\hline 
\multicolumn{1}{|c|}{\textbf{}} & 
\multicolumn{1}{c|}{\textbf{N}} &
\multicolumn{1}{c|}{\textbf{Minimum}} &
\multicolumn{1}{c|}{\textbf{Maximum}} &
\multicolumn{1}{c|}{\textbf{Mean}} &
\multicolumn{1}{c|}{\textbf{{\footnotesize Std. Deviation}}} \\
\hline 
\endhead
\hline \multicolumn{6}{|r|}{{Continued on next page}} \\ \hline
\endfoot
\hline \hline
\endlastfoot
\textbf{Advice Was Helpful} & 7 & 4 & 5 & 4.14 & 0.378 \\
\hline
\textbf{Options Were Easy to Follow} & 7 & 4 & 5 & 4.14 & 0.378 \\
\hline
\textbf{Found Advice Wanted} & 7  & 3 & 4 & 3.86 & 0.378 \\
\hline
\textbf{Would Have on Desktop or Phone} & 7 & 2 & 5 & 3.57 & 0.976 \\
\hline
\textbf{Helpful to Change Behaviour} & 7 & 2 & 5 & 3.14 & 1.069 \\
\hline
\end{longtable}
\end{center}
The table shows the the \texttt{Sample Size}, \texttt{Minimum} score, \texttt{Maximum} score, \texttt{Mean} and \texttt{Standard Deviation} for each of the five questions on the questionnaire.

Unfortunately, none of the results  are statistically significant.  However, the results indicate that using the system reduces the time to get advice about a particular task compared to just reading it from a list, and that the system was found to be useful to the users.

\section{Conclusion and Future Work}
\label{conclusions}

We have described EcoHomeHelper and how it has been useful in solving the problem of organizing and presenting large quantities of data to the user in an efficient manner.  We have looked at specific code which was important in the execution of the application. Further details of the project are found at http://homepage.cs.latrobe.edu.au/sloke/projects/EcoHomeHelper.html, with the full knowledge base of advice facts, and their sources.

While this writer concedes that the results were not statistically significant, it is more due to a lack of power or sample size.  A larger sample size may have yielded different results.  Another aspect which could have given significantly different results, is if readers were given texts to search through and find the relevant advice relating to the scenarios.  In short, this study can be considered as a pilot study and further future research into this area is needed.

There are positives though.  The application can be used as a web based application that users could access.  Many councils offer advice on which residents can use to reduce their GHG emissions.  The user is still required to read from a web page and cipher through unnecessary information.  EcoHomeHelper can be placed on the web using Prolog Server Pages \cite{ben08}.  It functions in a similar way to PHP.  An added improvement while acting as a web application is that links could be added to the advice.  If the user was interested in buying a fridge, a link could take the user to a page displaying makes and models of fridges and their energy rating, or if there were environmentally friendly.

EcoHomeHelper could be developed into an application for a mobile phone.  Users could access advice on GHG emissions from their phone.  While not statistically significant, results from this work suggest that younger people are more comfortable with using technology in their daily life compared to older people.

Finally, the mobile phone could not only give advice, but could alert the user when he or she enters a shop that sells appliances such as fridges or TVs.  This does not differ greatly in concept from the GPS system.  When the user enters a shop such as ``The Good Guys'', the phone will sound an alert and bring up a menu.  The user then can make a number of choices or exit the application.

\bibliographystyle{plain}
\bibliography{MattDonatoResearchPaper}

\end{document}